\documentclass[11pt]{article}
\usepackage{graphicx}
\usepackage{pgfplots}

\usepackage[colorlinks=true,citecolor=blue]{hyperref}
\usepackage{natbib}
\usepackage{graphicx}
\usepackage{amsfonts}
\usepackage{amsmath}
\usepackage{amssymb}
\usepackage{url}
\usepackage{fancyhdr}
\usepackage{indentfirst}
\usepackage{enumerate}
\usepackage{titlesec}
\usepackage{amsthm}
\usepackage{subcaption}
\usepackage{dsfont}
\usepackage[misc]{ifsym}

\theoremstyle{definition}

\theoremstyle{plain}

\theoremstyle{remark}

\usepackage{cases}

\theoremstyle{definition}

\def\P{\mathbb{P}}
\def\p{\mathbb{P}}
\def\E{\mathbb{E}}

\def\R{\mathbb{R}}

\def\var{\mathrm{var}}

\DeclareMathOperator{\sign}{sgn}

\def\d{\mathrm{d}}

\newcommand{\VaR}{\mathrm{VaR}}
\newcommand{\ES}{\mathrm{ES}}

\renewcommand{\(}{\left(}
\renewcommand{\)}{\right)}

\DeclareMathOperator*{\argmin}{arg\,min}

\def\laweq{\buildrel \d \over =}

\usepackage[onehalfspacing]{setspace}

\usepackage{bm}
\usepackage{tikz-qtree}
\usepackage{tikz}

\def\id{\mathds{1}}

\title{Infinite-mean models in risk management: Discussions and recent advances}

\author{Yuyu Chen\thanks{Department of Economics, University of Melbourne,  Australia. \Letter~{\scriptsize\url{yuyu.chen@unimelb.edu.au}}} 
\and Ruodu Wang\thanks{Department of Statistics and Actuarial Science, University of Waterloo,  Canada. \Letter~{\scriptsize\url{wang@uwaterloo.ca}}} 
}
\date{\today}

\begin{document}

	\maketitle
	\begin{abstract}  
 	In statistical analysis, many classic results require  the assumption that models have finite mean or variance, including the most standard versions of the laws of large numbers and the central limit theorems. 
 Such an assumption may not be completely innocent, and it may not be appropriate for datasets with heavy tails (e.g., catastrophic losses), relevant to financial risk management. In this paper, we discuss the importance of infinite-mean models in economics, finance, and related fields, with recent results and examples. We emphasize that many results or intuitions that hold for finite-mean models turn out to fail or even flip for infinite-mean models. Due to the breakdown of standard thinking for infinite-mean models, we argue that if the possibility of using infinite-mean models cannot be excluded, great caution should be taken when applying classic methods that are usually designed for finite-mean cases in finance and insurance.
	
% \textbf{JEL classification}: C02; C62; G22. 

	\end{abstract}

\section{Introduction}\label{sec:1}

% 	[I have three keywords in the title (they can be changed): fascination means interesting new results and phenomenon (e.g., something that is flipped from the finite-mean case);
% 	precaution means that some methods need to be applied with great care (e.g., statistical methods, estimators, confidence intervals, VaR); 
% 	prohibition means that something cannot be used (e.g., central limit theorem, coherent risk measures). We can center our discussions around these three types of results or situations. ]

The purpose of this article is to review and discuss recent advances in infinite-mean models and their intriguing behavior and implications in risk management and statistics. The discussion is by no means exhaustive, but rather it reflects the subjective choices of the authors and their  limitation of knowledge. 

Let us first set straight a simple, but perhaps important and puzzling, question. 
Whereas whether the world is finite or not may be philosophical or agnostic,  it is agreed by most people that the financial world is finite, because the total amount of money in the world is finite.
Why are we interested in a model with infinite mean, when the risk variables that we try to model  are eventually bounded?

The short answer is that these infinite-mean models  fit statistically better the reality in many relevant situations, and they provide useful insights on risk management problems, which are sometimes as useful as finite-mean models, but very different in nature, as we will discuss.

Important to infinite-mean models is the principle of power law. 
 Power-law models have been extensively employed to study various heavy-tailed phenomena such as population incomes, city sizes, word frequencies, stock returns, and financial losses. They are also made popular to the public via its connection to the Black Swan of \cite{T07}.  \cite{andriani2007beyond} and \cite{clauset2009power} discussed   the applications and empirical evidence of power-law models. Although models with finite moments are commonly assumed in many statistical analyses, in the context of finance and insurance, power-law models with divergent second or even first moments are ubiquitous.

 Since empirical observations are finite by nature, sample moments of all orders are always finite. If sample moments converge for large numbers of sample sizes, it is useful to take the ``stabilized" values as approximations of the true moments due to the law of large numbers. In view of the finiteness of sample moments,  models with infinite moments may seem problematic for applications. However, as \cite{mandelbrot2013fractals} emphasized, the concept of infinite moments is not ``improper" as sample moments do not always converge or may tend to infinity as the sample size increases. Therefore, the finiteness of sample moments cannot be used to exclude the possibility of using models with infinite moments (see also p.~88--92 of \cite{mandelbrot2013fractals}). In practice, of course, one can avoid infinite moments by either putting bounds to infinite-mean models or replacing their tail parts with lighter tails. As a cost, these corrections may give varying or even misleading results.

Infinite-mean models are by no means strange or pathological. For an inexperienced reader, let us consider a uniformly distributed random variable $U$ on $[0,1]$, or a normally distributed random variable $X$ with mean $0$ and variance $1$, both very common in any probabilistic or statistical analysis. What may not be obvious, although mathematically simple to verify, is that $1/U$ and $1/X$ have infinite mean, although they are very natural objects. (Note that $X$ takes the value $0$ with probability $0$, so the reciprocal $1/X$ is well-defined.)

 Among models with infinite moments, those with infinite means are particularly useful in risk management, especially for modeling catastrophic risks and operational risks.  
 Section \ref{sec:3} contains a summary of various examples of infinite-mean models in risk management and related areas.  In this paper, we collect some recent results and phenomena of infinite-mean models across different research domains.  It turns out that the assumption of having a finite mean is crucial for many classic results to hold; many of these results are typically flipped for infinite-mean models. We also emphasize that several commonly used methods in statistics, decision analysis, and risk management are not useful for infinite-mean models anymore. For instance, the classic central limit theorem requires a finite second moment and sample moments will not converge for infinite-mean models.  Therefore, if infinite-mean models are the outcomes of careful statistical analysis, standard methods should be used with considerable caution or possibly avoided.

 The rest of the paper is organized as below. Section \ref{sec:2} gives several examples of power-law distributions. 
 Section \ref{sec:3} exemplified the occurrences of infinite-mean models in many studies. 
 Standard estimation methods of power-law models are briefly described in Section \ref{sec:EVT}. In Section \ref{sec:5}, we point out several  statistical, decision, and risk models that may fail for infinite-mean models. Sections \ref{sec:6}--\ref{sec:other} discuss the applications of infinite-mean models in risk aggregation,  optimization and risk sharing, multiple hypothesis testing, optimal bundling, firm growth analysis, and linear estimators, respectively. Section \ref{sec:conclusion} concludes the paper.

\section{Definitions and properties}\label{sec:2}

% [Basic stuff; Pareto distributions; Cauchy;  I suggest that we do not explain too complicated models. When needed, just say ``this result can be generalized to ... see ..."]

Many useful models applied to heavy-tailed phenomena fall into the class of power-law distributions. A random variable $X$ is said to follow a \emph{power-law distribution} (or a regularly varying distribution) with tail parameter $\alpha>0$, if
$$\P(X>x)= \frac{L(x)}{x^{\alpha}},$$
where $L$ is a \emph{slowly varying} function, that is, $L(tx)/L(x)\rightarrow 1$ as $x\rightarrow \infty$ for all $t>0$.
% Here, $g(x)\sim h(x)$ as $x\rightarrow \infty$ means $\lim_{x\rightarrow \infty}g(x)/h(x)=1$. 
We say a power-law distribution is \emph{extremely heavy-tailed} if $\alpha\le 1$. A distribution is said to be an \emph{infinite-mean distribution/model} if its absolute moment of order 1 is infinite (i.e., a random variable $X$ follows an infinite-mean distribution if $\E[|X|]=\infty$). We are particularly interested in those models for $X$ with $\E[X_+]=\infty$, where $X_+=\max\{X,0\}$, for their financial interpretation. We present below some important examples of power-law distributions, all of which are infinite-mean models if their tail parameters are no greater than 1.

The most popular power-law model is arguably the \emph{Pareto distribution}, with its distribution function defined as
 \begin{align*}%\label{eq:PD}
  \p(X\le x) = 1 -\left(\frac{\theta}{x}\right)^{\alpha},~~x\ge \theta,
  \end{align*}
where $\alpha>0$ is the tail parameter and $\theta>0$ is the scale parameter. The $n$th moment of $X$ is infinite if and only if the tail parameter $\alpha $ is  in $ (0,n]$;  in particular, $\E[X]=\infty$
 if $\alpha \in (0,1]$. This parametrization of the Pareto distribution is known as the Pareto distribution of Type I; see \cite{A15} for other parametrizations of Pareto distributions and the related distributional properties.

 An important distribution from the Extreme Value Theory (EVT) is the generalized Pareto distribution (GPD). The \emph{generalized Pareto distribution} with parameters $\xi\in \R$ and  $\beta>0$ is defined as 
 \begin{equation}\label{eq:GPD}
     \p(X\le x)=
    \begin{cases}
     1-\(1+\xi\frac{x}{\beta}\)^{-1/\xi},&\mbox{~~if $\xi\neq 0$}, \\
    e^{-x/\beta}, & \mbox{~~if $\xi= 0$},
    \end{cases}
\end{equation}
where $x\in[0,\infty)$ if $\xi\ge 0$ and $x\in[0,-\beta/\xi)$ if $\xi< 0$.     
   If $\xi>0$, the generalized Pareto distribution is a power-law distribution with a tail parameter being $1/\xi$. If $\xi\ge 1$, then the generalized Pareto distribution has infinite mean. By the Pickands-Balkema-de Haan Theorem \citep{BD74, P75},  the generalized Pareto distributions are the only possible non-degenerate limiting distributions of the excess of random variables beyond a high threshold.

Another important power-law distribution in the EVT literature is the \emph{Fr\'echet distribution} with the distribution function given by
$$\p(X\le x)=\exp(-x^{-\alpha}), ~~x>0,$$
where $\alpha>0$ is the tail parameter.
If $\alpha\le 1$, then $\E[X]=\infty$. By the Fisher-Tippett-Gnedenko Theorem, the Fr\'echet distribution (with location-scale transforms) is one of the three possible non-degenerate limiting 
 distributions of the maximum of iid random variables (the limiting distributions are referred to as the extreme value distributions). In particular, if $X_1,X_2,\dots$ are iid regularly varying random variables, then there exist sequences of constants $\{d_n\}$ and $\{c_n\}$ where $c_n>0$ such that $(\max\{X_1,\dots,X_n\}-d_n)/c_n$ converges to the Fr\'echet distribution (up to location-scale transforms).

 Two other examples of extremely heavy-tailed power-law models are the \emph{Cauchy distribution} and the \emph{L\'{e}vy distribution}. The standard Cauchy cumulative distribution function is  
 $$\p(X\le x)=\frac{1}{\pi}\arctan(x)+\frac{1}{2},~~x\in\R,$$
 and the standard L\'{e}vy cumulative distribution function is
 $$\p(X\le x)=2-2\Phi\(\frac{1}{\sqrt{x}}\),~~x\ge 0,$$
 where $\Phi$ is the cumulative distribution function of the standard normal distribution. 
 The tail indices of the Cauchy distribution and the L\'{e}vy distribution are $1$ and $0.5$, respectively. The expectation of the Cauchy distribution does not exist and that of the L\'{e}vy distribution is $\infty$.

 The Cauchy distribution is also well-known via its density function, given by
 $$
f(x)=  \frac{1}{\pi(1+x^2)},~~x\in \R.
 $$
This function is also known as ``the witch of Agnesi", getting the name from a book of Maria Gaetana Agnesi published in 1748.
In any college-level probability courses, the Cauchy distribution is the first example of an infinite-mean distribution, for which the laws  of large numbers fail. 

 Both the Cauchy and the L\'{e}vy distributions are special cases of a large class of power-law distributions called stable distributions. The characteristic function of a \emph{stable distribution}
with stability parameter $\alpha \in (0,2)$,
 skewness parameter $\beta\in[-1,1]$, scale parameter $\sigma=1$ and shift parameter $\mu=0$,
  is 
\begin{equation*}
    \varphi(\theta) =\begin{cases}
    \exp\left(-|\theta|^\alpha (1-i\beta\sign(\theta)\tan\frac{\pi\alpha}{2})\right) \mbox{~~~~if $\alpha\neq 1$},\\
    \exp\left(-|\theta|(1+i\beta\frac{2}{\pi}\sign(\theta)\log|\theta|)\right) \mbox{~~~~if $\alpha= 1$},
    \end{cases}
    \mbox{$\theta \in \R$,}
\end{equation*}
where $\sign(\cdot) $ is the sign function.
 As the name suggests, a linear combination of iid stable random variables follows the same distribution of the iid random variables, up to location-scale transforms. Another important property of stable distributions is that they are the only possible limiting distributions of sums of iid random variables; see \cite{S17} and \cite{N20}  for properties of stable distributions.  Stable distributions are usually defined via their characteristic functions because they generally do not have analytical cumulative distribution functions. The few examples with analytical cumulative distribution functions include the normal distribution, the Cauchy distribution ($\alpha=1$, $\beta=0$), and the L\'{e}vy distribution ($\alpha=1/2$, $\beta=1$). Stable distributions with $\alpha<1$ and $\beta=1$ are defined on the positive axis and have infinite mean. A stable distribution has infinite absolute expectation if its tail parameter is less than or equal to 1.

\section{Empirical evidence in finance and insurance}\label{sec:3}

% [Something similar to the review we did. It would be nice if it has more content or details than our other paper. We should look at other references as well.]

Compared with distributions with infinite second or higher moments, statistical models with infinite mean are less common in finance and insurance literature and thus need more motivation. 
Recall that 
financial asset return data typically show features of finite mean and infinite 4th or 5th moments; see the summary of \cite{C01}. Moreover, \cite{mandelbrot2013fractals} provided many examples of infinite second moments. 
Nevertheless, there are still many contexts where infinite-mean models are useful.
In the following discussion, we delve into examples in the literature leading to extremely heavy-tailed power-law models. 
Our list is only for a simple illustration, and it is by no means exhaustive. 

\subsection*{Catastrophic losses}

With standard seismic theory,  \cite{IJW09} illustrated that the tail indices $\alpha$ of earthquake losses lie in the range $[0.6,1.5]$. The tail indices $\alpha$ for some wind catastrophic losses are around $0.7$, as shown by \cite{rizzo2009new}. \cite{CEW24b} reported that the tail indices $\alpha$ of the wildfire suppression costs in Alberta, Canada are around $0.85$. The tail indices of losses caused by nuclear power accidents are observed in the range of  $[0.6,0.7]$ by \cite{hofert2012statistical}; similar observations were made by \cite{sornette2013exploring}.

\subsection*{Operational losses}

By analyzing data collected by the Basel Committee on Banking Supervision, \cite{moscadelli2004modelling} reported the tail indices $\alpha$ of operational losses in $8$ different business lines to lie in the range $[0.7,1.2]$. Notably, $6$ out of the $8$ tail indices are less than $1$, with $2$ being significantly less than $1$ at a $95\%$ confidence level. See \cite{NEC06} for a detailed discussion on the risk management consequences of infinite-mean operational losses. Cyber risk losses exhibit tail indices $\alpha\in[0.6, 0.7]$, as explored by \cite{EW19} and \cite{ES20}.

\subsection*{Large insurance losses}

    Several significant commercial property losses collected from two Lloyd’s syndicates have tail indices $\alpha$ considerably less than $1$; see \cite{BC14}. Analyzed by \cite{beirlant1999tail}, some fire losses collected by the reinsurance broker AON Re Belgium have tail indices $\alpha$ approximately equal to $1$. \cite{CEW24b} showed that the tail parameter $\alpha$ for the marine losses of a French private insurer is around $0.9$. Most  major damage insurance losses in a standard Swiss Solvency Test document (\citet[p.~110]{FINMA2021}) 
  have default tail parameter $\alpha $ in the range $[1,2]$, with  $\alpha=1$ attained by some aircraft insurance.

\subsection*{Financial data}

According to \cite{silverberg2007size},  the tail indices of financial returns from some
technological innovations are less than $1$.  The tail part of cost overruns in information technology projects can have tail indices $\alpha\le 1$, as discussed by \cite{flyvbjerg2022empirical}. The tail indices of detected fraud sizes in firms were estimated to be $0.65$ by \cite{cheynel2022fraud} who also proposed a theoretical mechanism to explain the heavy-tailed phenomenon. \cite{gabaix2003theory} showed that the market value of the managed
assets in top mutual funds has tail indices around $1$.

\subsection*{Other infinite-mean power laws}

 Word frequencies, city sizes, and firm sizes follow Zipf's law ($\alpha\approx 1$); see \cite{Z49}, \cite{gabaix1999zipf}, and \cite{axtell2001zipf}. Interconnected agglomerations of firms in Italy have a tail index around $1$; see \cite{andriani2007beyond}. Casualties in major earthquakes and pandemics modelled by Pareto distributions also have tail indices less than $1$; see \cite{clark2013note} and \cite{cirillo2020tail}. War sizes and intensities fit well with infinite-mean power law models (\cite{clauset2009power} and \cite{cederman2003modeling}). Both peak gamma-ray intensity of solar flares and the number of religious followers have tail indices around $0.8$ (\cite{clauset2009power}).

\section{Identifying infinite mean using extreme value theory}\label{sec:EVT}

% [Explain the basic inference of infinite-mean models: how do we estimate the tail parameter from data. Textbook stuff.]

Since data are finite but our interested models have no finite mean, it is important to verify whether a given dataset fits with a model with infinite mean. This is typically done by estimating the tail parameter. It needs to be emphasized that if the possibility of infinite-mean models cannot be excluded, one should consider using statistical methods based on order statistics rather than those based on moments. Below we only present some basic estimation methods for the tail parameter and we refer readers to \cite{EKM97}, \cite{DF06}, and \cite{RT07} for comprehensive discussions and more advanced methods.

Standard methods to estimate the tail parameter of power-law distributed data include the classic Hill method and the GPD method from the EVT (see \cite{MFE15} and \cite{EKM97}). Another estimation method is the log-log rank-size estimation (see \cite{ibragimov2015heavy} and \cite{gabaix2011rank}). In this section, we briefly describe the Hill method and the GPD method; we refer to \cite{EKM97} and \cite{MFE15} for more details.

\begin{figure}[h]
\centering
\includegraphics[height=7cm, trim={0 0 0 20},clip]{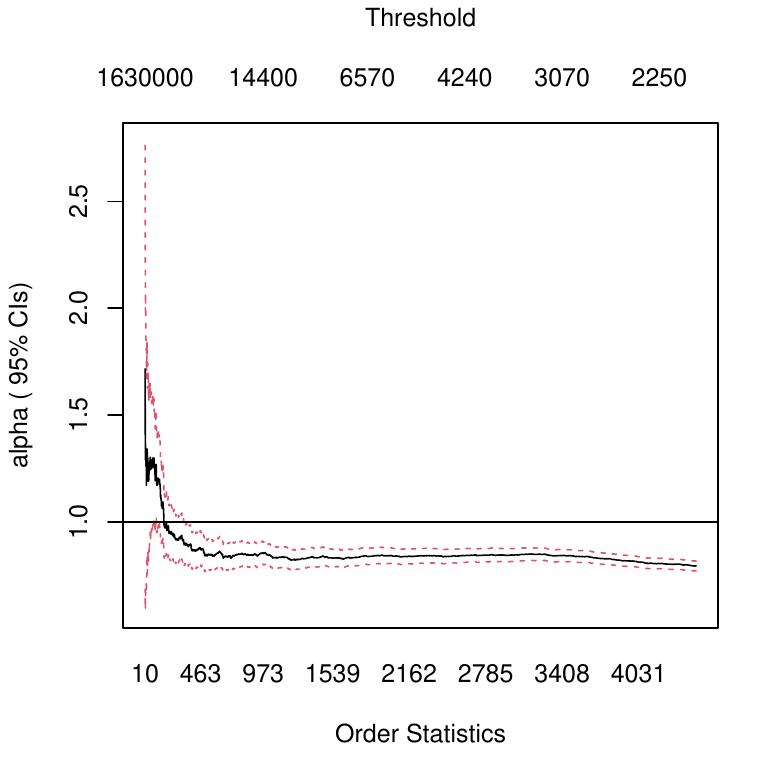}
\caption{The Hill plot for wildfire suppression costs: The Hill estimates are plotted as black curves with the $95\%$ confidence intervals being red curves. Adapted 
     from \cite{CEW24b}.
     % ``Risk exchange under infinite-mean Pareto models," by Y. Chen, P. Embrechts, and R. Wang, 2024, \emph{arXiv:2403.20171}.
     }
\label{f1}
\end{figure}

Let $X_1,\dots,X_n$ be iid non–degenerate random variables following a power-law distribution with tail parameter $\alpha>0$. Denote the order statistics of $X_1,\dots,X_n$ by $X_{n,n},\dots,X_{1,n}$, from the smallest to the largest (i.e., $X_{n,n}\le\dots\le X_{1,n}$). The Hill estimator of $\alpha$, denoted by $\hat \alpha_{k,n}$, is defined as
\begin{equation}\label{eq:Hill}
\hat \alpha_{k,n}=\(\frac{1}{k}\sum_{j=1}^k\log X_{j,n}-\log X_{k,n}\)^{-1},~~ 2\le k\le n.
\end{equation}
Thus the Hill estimator is based on the $k$ largest order statistics of an iid sample; some variations of the Hill estimator replace $k$ in \eqref{eq:Hill} by $k-1$. The value of $k$ is generally small and is chosen from the Hill plot which gives the Hill estimates $\hat \alpha_{k,n}$ for different values of $k$. If there is a stable region in the Hill plot where the Hill estimates are close for different $k$, then the Hill estimates are obtained from the stable region. In practice, one can roughly choose $k$ such that $k/n\in[0.01,0.05]$ (see \cite{MFE15}). We take the Hill plot for wildfire suppression costs in Alberta, Canada as an example (Figure 1 (b) of \cite{CEW24b}); see Figure \ref{f1}. By choosing a threshold around the top $5\%$ order statistics of the sample, we can see a stable region after the threshold in Figure \ref{f1} and the Hill estimate is around 0.847.

The Hill method works well for data whose tail distribution is close to a perfect power function but it can also be misleading in some cases. For instance, the Hill plot for AT\&T stock data in Figure 5.8 of \cite{MFE15} does not have a stable region. It is also known that the Hill method can be sensitive to the serial dependence in the data. Asymptotic properties of the Hill estimator can be found in  \cite{EKM97}.

The GPD method assumes that the excess distribution of $X_1,\dots,X_n$ over some high threshold $u$ is a GPD distribution \eqref{eq:GPD}. The tail parameter of the GPD distribution can be estimated by the maximum likelihood method and the probability-weighted moments method (see \cite{EKM97}).

However, there is a common method for GPD distribution that does not work for the infinite-mean case. To diagnose whether a GPD distribution is a good fit for the excess distribution with threshold $u$, the sample mean excess function is often used. For $v>0$, a sample mean excess function is defined as  $$e_n(v)=\frac{\sum_{i=1}^n(X_i-v)\id_{\{X_i>v\}}}{\sum_{i=1}^n\id_{\{X_i>v\}}}.$$
 If a GPD distribution is a good fit, then $e_n(v)$ 
 should be ``linear" for $v\ge u$ given large $n$; this is because the mean excess function $e(v)=\E[X-v|X>v]$, $v>0$, is linear in $v$ if $X$ is a GPD random variable. Such a graphic diagnosis will not work if the GPD distribution has infinite mean. \cite{klar2023pareto} proposed a new diagnosis method without moment restrictions.

\section{Decision and statistical models that need extra care}\label{sec:5}

\subsection{The St.~Petersburg paradox}

The most famous example of an infinite-mean model in decision making is perhaps  the St.~Petersburg paradox. Although it is well known, we briefly discuss it here before moving to a more formal analysis.

Consider a lottery that yields a payoff of $2^n$ dollars with probability $2^{-n}$, for $n=1,\dots,\infty$. This lottery can be achieved easily by throwing a fair coin many times until a tail has been observed, and a payoff of $2^n$ is made to the player if it takes $n$ rounds to reach a tail (i.e.,~there are $n-1$ heads before this round).
It is easy to observe that the expected payoff is $\sum_{n=1}^\infty 2^{-n} 2^n =\infty$. 
Since this payoff has an infinite mean, for a risk-neutral player the fair price of the lottery should be $\infty$; however, few people are willing to pay more than 20 dollars to play this game, thus the paradox. 

Daniel Bernoulli's answer in 1738 (translated version in \cite{B54}) to this paradox is to use a logarithm utility function to model the decision maker, which leads to a finite expected utility of the lottery. However, it is easy to design another lottery with an infinite expected utility for the logarithm utility function.
This creates the same trouble as in the original paradox: In theory, 
the player should prefer such a lottery, and even a tiny fraction of such a lottery (i.e., multiplying the payoff by 
 a very small $\epsilon>0$), to an arbitrarily large constant payoff, which is clearly paradoxical. 
There are many other explanations for the paradox, and we omit them. Here, we simply would like to show that an infinite-mean model can lead to controversial situations, even considering a simple lottery.

 \subsection{Stochastic dominance}

Notions of stochastic dominance are commonly used in decision analysis to rank risks; see \cite{levy1992stochastic,L16} for wide applications of stochastic dominance. Two commonly used stochastic dominance relations are stochastic order and convex order. Let $X$ and $Y$ be two random variables. Then
 \begin{enumerate}[(i)]
     \item $X$ is less than $Y$ in \emph{stochastic order}, denoted by $X\le_{\rm st}Y$, if $\p(X> x)\le \p(Y> x)$ for all $x\in\R$; 
     \item $X$ is less than $Y$ in \emph{convex order}, denoted by $X\le_{\rm cx}Y$, if $\E[u(X)]\le \E[u(Y)]$ for all convex functions $u$ such that the two expectations exist.
 \end{enumerate}
 
 These two notions of stochastic dominance have distinct meanings and provide very different implications in decision makings. In general, $X\le_{\rm st}Y$ means that $Y$ is ``larger" than $X$. Indeed, there exists another random variable $X'$ such that $X'$ follows the same distribution of $X$ and $X'\le Y$ holds almost surely (e.g., Theorem 1.A.1 of \cite{SS07}). On the other hand,  $X\le_{\rm cx}Y$ means that $Y$ is more ``spread-out" than $X$; for instance, it implies $\var(X)\le \var(Y)$, but generally stronger than that. See \cite{SS07} and \cite{MS02} for comprehensive treatment of these and other stochastic dominance relations. 
 
When comparing infinite-mean risks, $\le_{\rm st}$ order always works whereas $\le_{\rm cx}$ may fail because the expectation of any increasing convex transform of a positive infinite-mean random variable is infinity. Both dominance relations work well for finite-mean models but $\le_{\rm st}$ order is not very useful in comparing risks with the same mean; e.g.,  in portfolio selection, it is common that different portfolios have the same mean. If $\E[X]=\E[Y]< \infty$ and $X\le_{\rm st}Y$, then $X$ and $Y$ have the same distribution (Theorem 1.2.9 of \cite{MS02}). Stochastic order $\le_{\rm st}$ provides very strong implications in making decisions as explained below.

 \subsection{Risk measures and expected utilities}

 We discuss several decision models that can or cannot be applied to infinite-mean models. A \emph{preference} is a function $\rho: \mathcal X_{\rho}\rightarrow \overline \R:= [-\infty,\infty]$, where the domain $\mathcal X_\rho$ is a set of random variables representing risks. For our discussion, we use the term preference in a very broad sense; 
  it includes risk measures in the sense of \cite{ADEH99} and \cite{FS16} and 
preference models such as the expected utility. For two random losses $X$ and $Y$, an agent equipped with a preference model $\rho$ prefers $X$ over $Y$ if $\rho(X)\le \rho(Y)$; for random profits $X$ and $Y$, the agent prefers $X$ over $Y$ if $\rho(Y)\le \rho(X)$. For the rest of this section, assume that all random variables represent losses.  

Most commonly used preference models for comparing risks have the following two properties. We write $ X\laweq  Y$ if $ X$ and $ Y$ have the same distribution.
 \begin{enumerate}[(a)]
 \item Law invariance: $\rho(X) = \rho(Y)$ for $X,Y\in\mathcal X_\rho$ if $X\laweq Y$.
 \item Monotonicity: $\rho(X) \le \rho(Y)$ for $X,Y\in\mathcal X_\rho$ if $X\le_{\rm st} Y$. 
%  \item Mild monotonicity: $\rho$ is weakly monotone and 
% $\rho(X)< \rho(Y)$ for $X,Y\in\mathcal X_\rho$ if $\p(X<Y)=1$.
\end{enumerate} 
Decision models that do not respect (a) are often regarded as models for decisions under ambiguity. We do not discuss them and interested readers may refer to \cite{W10}.
We will call a preference model satisfying (a) and (b) a monotone risk preference. 

 If random variables $X$ and $Y$ represent losses, an agent equipped with a monotone risk preference will prefer $X$  over $Y$ as soon as $X\le_{\rm st}Y$. We present below some examples of monotone preference models.
 
 The first examples are Value-at-Risk (VaR) and Expected Shortfall (ES), the two regulatory risk measures in the insurance and finance sectors.
  For a random variable $X$ and $p\in (0,1)$, VaR is defined as
  \begin{equation*}%\label{VaR}
  \VaR_{p}(X)=\inf\{t\in\R:\p(X\le t)\geq p\},
  \end{equation*}
and ES is defined as
$$\ES_{p}(X)=\frac{1}{1-p}\int_{p}^{1}\VaR_{u}(X)\mathrm{d}u.$$
%If $F_X$ is continuous, then $\ES_{p}(X)=\E[X|X\ge \VaR_p(X)]$. 
 If $X$ has infinite mean, $\VaR_p(X)$ is always finite whereas $\ES_{p}(X)$ can be infinite and is thus not very useful for infinite-mean models.  Many useful risk measures for infinite-mean models are in the class of distortion risk measures such as VaR and Range Value-at-Risk proposed by \cite{cont2010robustness} and studied by \cite{ELW18}; distortion risk measures coincide with dual utilities of \cite{Y87} up to a sign change. By contrast, convex risk measures (\cite{follmer2002convex}), which have several nice properties,  will take infinite values when evaluating infinite-mean losses and hence are not useful (see \cite{filipovic2012canonical}).

The next example is the expected utility model. Let $u$ be an increasing   function, called a utility function, and $u(w)$ represents the utility of wealth $w$.  Since we are dealing with losses, we flip the sign of the expected utility model, by  using an increasing function $v$ (called a disutility function) given by $v(x)=-u(-x)$, and denote the expected utility model by $E_v$, that is,
$$
E_v(X) = \E[v(X)]= \E[-u(-X)].
% ~~~~~~X\in \X_{E_v}:=\{Y\in \X: \E[|v(Y)|]<\infty\},
$$
  An expected utility agent is said to be \emph{risk-averse} (in the sense of \cite{RS70}) if  $u$ is concave ($v$ is convex). Note that risk-averse expected utility may not be useful for comparing infinite-mean risks as $E_v$ on them can take infinite values due to the concavity of $u$. However, there are still some useful examples of expected utility models such as (i) the Markowitz utility function of \cite{M52b} ($u$ has a convex-concave structure on the loss side); (ii) the utility function in the cumulative prospect theory of \cite{TK92}
($u$ is convex below a reference point); (iii) agents with limited liability ($u$ is bounded from below).

Although risk-averse expected utility may be infinite for extremely heavy-tailed risks, it does not necessarily mean that there is no way to distinguish the risks for agents with concave utilities. 
 Instead of calculating their expected utility, an alternative way to compare risks is the ``overtaking criterion" used by optimal accumulation problems in economics; see \cite{V65}. %Let $c_1(t)$ and $c_2(t)$ be two consumption programs of an economy as a function of time $t$, and $w$ be a utility function of the consumption and time. The ``overtaking criterion" says that $c_1$ is better than $c_2$ if there exists some constant $T_0>0$ such that for all $T\ge T_0$, 
%$$\int_0^T w\(c_1(t),t\)-w\(c_2(t),t\)\mathrm{d} t> 0.$$
This ``overtaking criterion" is used to avoid the problem of infinite expected utility.
The key observation here is that, whenever $\E[v(X)]$ is well-defined, it can be rewritten as
$$
\E[v(X)]=\int_0^1 v(F_X^{-1}(p))\d p,
$$
where $F^{-1}_X$ is  the left quantile function of $X$, that is, $F^{-1}_X(p)=\VaR_p(X)$ for $p\in (0,1)$.
Applying this idea to compare infinite-mean losses, we propose the following formulation.
For two losses $X$ and $Y$,  we may say an agent with  model $E_v$ prefers $X$ over $Y$ if 
\begin{equation}\label{eq:overtaking}
\int_0^1 \left(v\(F^{-1}_X(p)\)-v\(F^{-1}_Y(p)\)\right)\mathrm{d} p< 0.\end{equation} 
In this form, since $X\le_{\rm st}Y$ implies $F^{-1}_X\le F^{-1}_Y$ on $(0,1)$,  we can conclude that
if $X\le_{\rm st}Y$ then $X$ is preferred to $Y$, even if 
$\E[v(X)]$ and $\E[v(Y)]$ are not well-defined.

 \subsection{Laws of large numbers and the central limit theorem}

 Some classical results for sums of independent random variables in statistical theory may fail for infinite-mean models.
 Although this observation is very simple, we would like to emphasize it as it can be overlooked by practitioners. 
 
 The first example is the law of large numbers. Let $X_1,X_2,\dots$ be a series of iid random variables and $S_n=\sum_{i=1}^nX_i$. By the strong law of large numbers, if  $\E[X_1]=\mu\in\R$, then $S_n/n\rightarrow \mu$ almost surely. That is, for a sample of iid observations, the sample mean will converge to the true expectation. However, if $\E[X_1]=\infty$, then  $S_n/n\rightarrow \infty$ almost surely (see, e.g., Theorem 2.4.5 of \cite{D19}). As the sample mean will explode for infinite-mean models, many applications based on the law of large numbers (e.g., Monte Carlo method) do not work.

 Another example is the central limit theorem which concerns the limiting distribution of $S_n$. Assume that $X_1,X_2,\dots$ are non–degenerate. If $\E[X_1^2]<\infty$, by the classic central limit theorem, the only limiting distribution of $S_n$ (after normalization) is the normal distribution. This result generally does not hold for heavy-tailed random variables as their second or even first moments are usually infinite.  In this case, the limiting distribution of $S_n$ (after normalization) for commonly used heavy-tailed random variables is a stable distribution; see, e.g., \cite{EKM97}.

\section{Some results on aggregating infinite-mean risks} \label{sec:6}

\subsection{Asymptotic super-additivity of VaR under regular variation}

 \label{sec:61}

 Suppose that $X_1,\dots,X_n$ are $n$ identically distributed risks faced by an insurer. It is well known that if  $(X_1,\dots,X_n)$ follows the multivariate normal distribution, VaR$_p$ is sub-additive for $p\in[0.5,1)$, i.e., 
 $$\VaR_p(X_1+\dots+X_n)\le \VaR_p(X_1)+\dots+\VaR_p(X_n).$$
This means that constructing a diversified portfolio helps to reduce the capital reserve level of the insurer. More generally, the sub-additivity of VaR holds if $(X_1,\dots,X_n)$ follows the elliptical distribution. For the above claims, see, e.g., Theorem 8.28 of \cite{MFE15}.

The above benefit from diversification however does not always exist outside the elliptical world. For instance, if $X_1,\dots,X_n$ are iid and follow a power-law/regularly varying distribution  with tail parameter $\alpha>0$, VaR is asymptotically super-additive for $\alpha< 1$, i.e., 
$$\lim_{p\rightarrow 1}\frac{\VaR_p(X_1+\dots+X_n)}{\VaR_p(X_1)+\dots+\VaR_p(X_n)}>1.$$
The above inequality flips for   $\alpha\ge 1$ and VaR is said to be asymptotically sub-additive. Therefore,
diversification may increase the insurer's risk if $X_1,\dots,X_n$ have infinite mean. The asymptotic properties of VaR also hold if $X_1,\dots,X_n$ are dependent, e.g., multivariate regularly varying $(X_1,\dots,X_n)$ (see 
\cite{ELW09} and \cite{mainik2010optimal}). We will extend the discussion of portfolio diversification
to a non-asymptotic sense in the following sections.

Besides the marginal behaviors of risks, their dependence structure is also crucial to the diversification effects. For instance,  if power-law risks $X_1,\dots,X_n$ are positively dependent via some Archimedean copula (see \cite{N06}), \cite{alink2004diversification} showed that for infinite-mean  (resp.~finite-mean) risks, a stronger dependence makes the portfolio $X_1+\dots+X_n$ ``smaller" (resp.~``larger") in the tail part. A similar observation for multivariate regularly varying risks is made in Chapter 13 of \cite{ruschendorf2013mathematical}.

An early observation of the detrimental effects of diversification was made by \cite{FM72}. It was shown that for infinite-mean stable models $X_1,\dots,X_n$, diversification makes the portfolio more ``spread-out", i.e., $(X_1+\dots+X_n)/n$ becomes more dispersive as $n$ increases. This was reasserted by \cite{ibragimov2005new} via probability inequalities, which will be discussed later.

\subsection{Stochastic dominance for extremely heavy-tailed risks}
Since the pioneering work of \cite{M52b}, diversification has been regarded as an efficient tool to reduce portfolio risks. In particular, if risks are iid and have a finite mean, diversification will make the portfolio less ``spread-out", as studied by \cite{S67} and \cite{FM72}. However, such results may not hold for infinite-mean risks, as we discussed in Section \ref{sec:61}. 

The comparison among diversified portfolios can be made in a stronger sense.
Recently, several comparisons of infinite-mean models have been made via stochastic dominance relations and probability inequalities. These results suggest that diversification will make a portfolio of infinite-mean risks ``larger" or more ``spread-out".

Let $X_{1},\dots,X_{n}$ be iid extremely heavy-tailed Pareto random variables. 
For a nonnegative vector $\(\theta_{1},\dots,\theta_n\)  $, \cite{CEW24a} showed  
\begin{equation}
 \label{eq:*}
      \theta X_{1}\le_{\rm st}\theta_{1}X_{1}+\dots+\theta_{n}X_{n} \mbox{~~~~where $\theta =\sum_{i=1}^n\theta_n $.}
\end{equation} 
Intuitively, the left-hand side of \eqref{eq:*} is a non-diversified portfolio on $X_1$ with total exposure  $\theta$ and the right-hand side of \eqref{eq:*} is a diversified portfolio with exposure $\theta_i$ allocated to $X_i$ for all $i=1,\dots,n$. If $X_1,\dots,X_n$ represent losses, \eqref{eq:*} suggests that any decision maker who prefers less loss will choose non-diversification, leading to a \emph{diversification penalty} (diversification will be preferred if $X_1,\dots,X_n$ represent profits). This result is ``unexpected" in the sense that (i) it holds in the strongest form of risk comparison, and (ii) it does not hold for finite-mean models; if $X_1,\dots,X_n$ have finite mean and $\min_i\theta_i>0$, \eqref{eq:*} holds if and only if $X_1,\dots,X_n$ are almost surely equal (Proposition 2 of \cite{CEW24a}).
Inequality \eqref{eq:*} also holds if $X_{1},\dots,X_{n}$ are weakly negatively associated super-Pareto random variables. The class of super-Pareto distributions includes special cases of the generalized Pareto, the Burr, the paralogistic, and the log-logistic distributions, all with infinite mean, as studied by \cite{CEW24a}. 

With the notion of majorization order, \cite{CHWZ24} generalized \eqref{eq:*}
by allowing comparison between different diversified portfolios of iid extremely heavy-tailed Pareto random variables. For two vectors $\(\theta_1,\dots,\theta_n\)$ and $\(\eta_1,\dots,\eta_n\)$ in $\R^n$,  $\(\theta_1,\dots,\theta_n\)$ is dominated by $\(\eta_1,\dots,\eta_n\)$ in \emph{majorization order} if 
$$
\sum^n_{i=1}\theta_i =\sum^n_{i=1}\eta_i \mbox{~~~and~~~}
         \sum^k_{i=1} \theta_{(i)} \ge \sum^k_{i=1} \eta_{(i)}\ \ \mbox{for}\ k\in [n-1],
$$
      where $\theta_{(i)}$ and $\eta_{(i)}$ represent the $i$th smallest order statistics of $\(\theta_1,\dots,\theta_n\)$ and $\(\eta_1,\dots,\eta_n\)$, respectively; see \cite{MOA11}. Let $\(\theta_{1},\dots,\theta_n\)$ and $\(\eta_1, \dots, \eta_n\)\in\R_+^n$ be the exposure vectors of two portfolios such that $\(\theta_1, \dots, \theta_n\)$ is smaller than $\(\eta_{1}, \dots, \eta_n\)$ in majorization order; the portfolio with exposure vector $\(\theta_1, \dots,\theta_n\)$ is more diversified than that with exposure vector $\(\eta_1, \dots,\eta_n\)$. \cite{CHWZ24} showed the following inequality 
\begin{align}
 \label{eq:main}
      \eta_1 X_1+\dots+\eta_n X_n \le_{\rm st} \theta_1 X_1 +\dots+\theta_n X_n.
\end{align}
Similar to \eqref{eq:*}, if $X_1,\dots,X_n$ represent losses, \eqref{eq:main} means more diversification leads to a worse portfolio in the sense that diversification makes the portfolio loss ``larger" (the observation is flipped if $X_1,\dots,X_n$ represent profits). Inequality \eqref{eq:main} also holds if $X_{1},\dots,X_{n}$ has a specific type of positive dependence structure modelled by Clayton copulas. By comparison, if $X_1,\dots,X_n$ are iid  with finite mean,  by Theorem 3.A.35 of \cite{SS07}, we have
\begin{equation}\label{eq:main-cx}
\theta_1 X_1 +\dots+\theta_n X_n\le_{\rm cx}   \eta_1 X_1+\dots+\eta_n X_n.
\end{equation}
Hence, diversification makes a portfolio of iid finite-mean losses less ``spread-out".

 Let $\(\theta_{1},\dots,\theta_n\)$ and $\(\eta_1, \dots, \eta_n\)$ be as before and $X_{1},\dots,X_{n}$ be iid extremely heavy-tailed stable random variables. By properties of stable distributions, \cite{ibragimov2005new} showed that,
\begin{equation}\label{eq:stable}
  \p(\eta_1 X_1+\dots+\eta_n X_n \ge x)\le \p(\theta_1 X_1 +\dots+\theta_n X_n\ge x)\mbox{~~for $x>0$},
\end{equation}
and the above inequality is flipped if $x< 0$. This result suggests that diversification makes the portfolio more ``spread-out", distinct from \eqref{eq:main}; similar observations can be found in \citet[p.~271]{FM72}. In particular, \eqref{eq:stable} implies that \eqref{eq:main} holds if $X_1,\dots,X_n$ are iid positive one-sided stable random variables with infinite mean. On the other hand, if $X_1,\dots,X_n$ are iid stable random variables with finite mean, the above inequalities will flip; that is, diversification will make the portfolio less ``spread-out". The result for finite-mean stable distributions continues to hold for symmetric log-concave distributions (\cite{P65}); a distribution with density is said to be \emph{log-concave} if its density is log-concave.  Inequality \eqref{eq:stable} also holds for convolutions of risks with joint $\alpha$-symmetric distributions (\cite{ibragimov2009portfolio}).
Similar probability inequalities for bounded stable random variables are studied by \cite{IW07}.

\subsection{The sum of infinite-mean risks can be a constant, but not unique}

The sum of infinite-mean risks can have strange behavior compared to finite-mean risks. Here we discuss one example that is relevant to risk management.  

Suppose that $n$ random variables $X_1,\dots,X_n$ follow the same distribution $F$, and they may be dependent. 
If  for some constant $C\in \R$ it holds that $\p(\sum_{i=1}^n X_i=n C)=1$, then we say that $C$ is a center of the distribution $F$. 
If $X_1,\dots,X_n$ have the same finite mean $\mu\in \R$, it is clear that $C=\mu$, since $\E[\sum_{i=1}^n X_i]=n\mu.$ However, if the mean of $X_1,\dots,X_n$ does not exist or is infinite, this becomes unclear.
A question stated by \cite{W15} was whether the center $C$ of $F$ is unique.
This question is answered by  \cite{PRWW19} for Cauchy distributions.
They showed that for every $n \geq 2$, there exist
$n$ standard Cauchy random variables adding up to a constant $nC$ if and only if
$|C|\le{\log (n-1)}/{\pi}.$ Hence, the center is not unique.

The existence of $C$ is formally studied through the notions of complete and joint mixability (\cite{WW11,WW16}), which  can be used as a building block for many 
optimization problems such as minimizing the variance or finding the worst-case value of risk measures in risk aggregation; see \cite{EPR13, EPRWB14}.
The relevance of this problem   to risk management is treated in 
 Chapter 8 of \cite{MFE15} 
 and its recent progresses are summarized in the monograph  \cite{RVB24}.

\section{Infinite-mean models in optimization and risk sharing}\label{sec:7}

We next discuss the implications of infinite-mean models in some optimization problems in risk management. 
 
\subsection{Portfolio optimization}

Suppose that an agent equipped with a risk preference model $\rho$ faces several iid profits $X_1,\dots,X_n$ and needs to maximize their preference:
$$\mbox{To maximize~~} \rho\(\sum_{i=1}^n\theta_iX_i\)\mbox{~~subject to~~} \theta_1,\dots,\theta_n>0 \mbox{~~and~~} \sum_{i=1}^n\theta_i=\theta,$$
where $\theta>0 $ represents the total budget of the agent. 
% to construct a portfolio $\sum_{i=1}^n\theta_iX_i$ to maximize their preference (e.g., expected utility), where $\theta_1,\dots,\theta_n>0$ and $\sum_{i=1}^n\theta_i$ is denoted by $\theta$. 
What can we say about the agent's optimal strategy? This question has been well studied for finite-mean models since the  work of \cite{M52a}, and the optimal strategy clearly depends on the agent's risk attitude. If the profits have finite mean,  the optimal strategy for risk-averse expected utility agents is full diversification (i.e., $(\theta_1,\dots,\theta_n)=(\theta/n,\dots,\theta/n)$) whereas that for risk-seeking expected utility agents is non-diversification (i.e., $(\theta_1,\dots,\theta_n)=(\theta,\dots,0)$); see \cite{S67}. Indeed, an expected utility agent prefers to diversify over iid finite-mean profits if and only if they are risk-averse (Theorem 4 of \cite{CHWZ24}).

For profits with infinite mean (e.g., profits from technology innovations; see \cite{silverberg2007size}), agents' risk attitudes may not affect their optimal decisions. Assume that the profits are iid extremely heavy-tailed Pareto random variables. In this case, as long as an agent is equipped with a monotone preference model, by \eqref{eq:main},  full diversification will maximize the agent's preference; see \cite{CHWZ24}. Consequently, for risk-averse expected utility agents facing iid profits that follow a Pareto distribution, the optimal strategy is always full diversification; this does not depend on whether the profits have finite mean or not.

Next, let us consider the case where the agent faces iid losses $X_1,\dots,X_n$. The optimization problem now becomes:
$$\mbox{To minimize~~} \rho\(\sum_{i=1}^n\theta_iX_i\)\mbox{~~subject to~~} \theta_1,\dots,\theta_n>0 \mbox{~~and~~} \sum_{i=1}^n\theta_i=\theta.$$
If $X_1,\dots,X_n$ have finite mean and the agent is equipped with a risk measure that is consistent with $\le_{\rm cx}$ order,\footnote{A risk measure $\rho$ is consistent with $\le_{\rm cx}$ order if for $X\le_{\rm cx}Y$, $\rho(X)\le\rho(Y)$.} by \eqref{eq:main-cx}, full diversification is preferred. If $X_1,\dots,X_n$ are extremely heavy-tailed Pareto risks,  the optimal decision is clearly no diversification, as for any monotone risk measure $\rho$, by \eqref{eq:*}, we always have  $\rho(\theta X_{1})\le \rho(\theta_{1}X_{1}+\dots+\theta_{n}X_{n})$. One important consequence is the super-additivity of VaR:
\begin{align*}\label{eq:supVaR}
    \VaR_p(\theta_1 X_{1})+\dots+\VaR_p(\theta_n X_{n})\le \VaR_p(\theta_{1}X_{1}+\dots+\theta_{n}X_{n})\mbox{~~for all $p\in(0,1)$}.
\end{align*}
See \cite{MFE15} for the consequences of the super-additivity of VaR. Note that this super-additivity result holds for all significance levels and is thus different from those on regularly varying distributions, which are in some asymptotic sense (\cite{ELW09}). For agents with concave utility functions,  although the expected utility model may take infinite value for extremely heavy-tailed risks, the agents can instead use the ``overtaking criterion" \eqref{eq:overtaking} which implies no diversification is still preferred.

Another interesting case is when the iid losses $X_1,\dots,X_n$ follow a symmetric stable distribution with tail parameter less than or equal to 1. If the agent is equipped with $\VaR_p$ where $p\in[0.5,1)$, \cite{ibragimov2009portfolio} showed that no diversification minimizes their risk measure. This result is flipped if $p\le 0.5$ or the tail parameter of the stable distribution is greater than 1; in either case, full diversification is optimal.

\subsection{Risk sharing}

Risk sharing is the allocation of risks among multiple parties and has been a fundamental tool for insurance companies to mitigate risks and build resilience. In the presence of heavy-tailed risks such as catastrophic losses, however, this standard method may fail and insurance companies may choose not to share risks or even not to participate in the market; see, e.g., \cite{IJW09}.

 \cite{CEW24b}  studied the effects of risk sharing in a risk exchange model. In the risk exchange economy, each agent is equipped with some position of an individual risk and the risks are iid.  The agents can exchange their risks with other agents by paying/charging premia. The goal of each agent is to minimize their risk assessment after risk exchange. 
 
 We fix some notations. Let  $\mathbf X=(X_1,\dots,X_n)$ be the risk vectors for $n$ agents. The initial exposure vector of agent $i$, denoted by $\mathbf a^i$, satisfies $\mathbf a^i \cdot\mathbf X=a_iX_i$ where $a_i>0$; that is, agent $i$ has an initial position $a_iX_i$. Let $\mathbf w^i\in\mathbb R_+^n$ be the exposure vector of agent $i$ on $\mathbf X$ after risk sharing and $\mathbf p=(p_1,\dots,p_n)\in \mathbb R_+^n$ be the premium vector for $\mathbf X$. The loss position of agent   $i\in[n]$ after risk sharing is 
 $$L_{i}(\mathbf w^i, \mathbf p)=\mathbf w^i \cdot\mathbf X - (\mathbf w^i-\mathbf a^i)\cdot\mathbf p.$$
 Agent $i\in[n]$ is equipped with a risk measure $\rho_i$ and a convex cost function $c_i$ with $c_i(0)=0$. The risk assessment of agent $i$ is
 $ \rho_i\(L_{i}(\mathbf w^i, \mathbf p)\) + c_i(\Vert \mathbf w^i \Vert -\Vert \mathbf a^i \Vert)$.
An \emph{equilibrium} of the market is 
 a tuple $\(\mathbf p^*,\mathbf w^{1*},\dots,\mathbf w^{n*}\) \in (\R_+^n)^{n+1}$ if the following  two conditions are satisfied.
 \begin{enumerate}[(a)]
  \item
  Individual optimality:
 \begin{equation*}\label{eq:opt-internal} \mathbf w^{i*}\in {\argmin_{\mathbf w^i\in\mathbb R_+^n}}\left \{ \rho_i\(L_{i}(\mathbf w^i, \mathbf p^*)\) + c_i(\Vert \mathbf w^i \Vert -\Vert \mathbf a^i \Vert)\right\},~~~ \mbox{for each }i\in[n].\end{equation*}
  \item
 Market clearance:
  \begin{equation*}\label{eq:clearance-internal}  \sum_{i=1}^{n}\mathbf w^{i*}=\sum_{i=1}^{n}\mathbf a^i.\end{equation*} 
  \end{enumerate}

If the risks are super-Pareto random variables (i.e., extremely heavy-tailed), even if there is some risk exchange at equilibrium, the agents will merely exchange their entire positions with each other; hence there is no risk sharing. A similar observation is made by \cite{IJW11} based on a different model. Moreover, the premia for all risks are the same. By contrast, if the risks have finite mean and the agents use ES as their risk measure, the agents will share risks proportionally according to their initial risk exposures in an equilibrium. The premia in this case are not the same as well. 

The situation becomes quite different if the
agents with initial infinite-mean risks can transfer their risks to some external agents. The external agents do not have any risks in the initial phase and can be regarded as institutional investors in the market model. Similarly, an equilibrium is attained in this setting if all agents, including the external agents, can minimize their risk assessments. If the external agents have
a stronger risk tolerance than the internal agents, both parties may benefit by transferring losses to the external agents at an equilibrium. Nevertheless, there is still no risk sharing, i.e., no agent will hold more than one risk at the equilibrium.

 In practice, due to the limited liabilities of insurers, one may bound each loss below some positive threshold. If the risks are positive random variables (e.g., Pareto risks), the bounded risks will have finite mean. In the model of \cite{CEW24b}, agents equipped with ES will prefer diversification over bounded risks. However, as studied in a different economic model of \cite{IJW09},  agents may not always prefer diversification of bounded risks. Their results show that insurers may fall into a ``nondiversification trap" for bounded extremely heavy-tailed losses; that is, even if insurers can benefit from high-level diversification, they may still prefer non-diversification as a low degree of diversification is harmful (the value of diversification is U-shaped in the number of risks).

% \section{Infinite-mean models and risk attitudes}

% [Our note (should be on arXiv by then)]

\section{Infinite-mean models in multiple hypothesis testing}\label{sec:9}

% [There are several papers, e.g., \cite{VW20}, \cite{W19}, \cite{LX20}, \cite{CLTW23}. Our new short paper also \cite{CWWZ24}.]

Infinite-mean risk models have recently been a popular tool in multiple hypothesis testing, a statistical problem that is relevant to many fields including risk management, such as testing the suitability of the credit rating of structured  finance products from ratings of different tranches. 

In the context of  multiple hypothesis testing, 
 a fundamental question is how to combine several p-values, or test statistics in general, into one p-value. 
A random variable $P$ is called a \emph{p-variable} if $\p(P\le p)\le p$ for all $p\in (0,1)$; realizations of p-variables are called p-values. For our discussions, it suffices to assume that the p-variables to merge are uniformly distributed on $(0,1)$, denoted by $U_1,\dots,U_n$.

Many useful methods to combine p-values are via the following generalized weighted averaging function
\begin{equation}\label{eq:mean}
  M^{\mathbf w}_{\phi}(p_1,\dots,p_n)
  :=
  \psi
  \left(
    w_1\phi(p_1)+\dots+w_n\phi(p_n)
  \right),
\end{equation}
where $\mathbf w=(w_1,\dots,w_n)\in [0,1]^n$ with $\sum_{i=1}^nw_i=1$, $\phi:[0,1]\to[-\infty,\infty]$ is a continuous strictly monotonic function, and $\psi$ is its inverse. Several previous methods in the class of \eqref{eq:mean} use quantile function $F^{-1}$ or $F^{-1}(1-\cdot)$ as $\phi$ to transform p-values, where $F$ is a cumulative distribution function; that is, transformations $\phi(U_1),\dots,\phi(U_n)$ follow distribution $F$. Some examples are  
\begin{enumerate}[(i)]
\item Fisher's combination method (\cite{F48}): $\phi(x)=-\log(x)$ for $x\in[ 0,1]$, and $\mathbf w=(1/n,\dots,1/n)$;
\item the harmonic mean method (\cite{W19}): $\phi(x)=1/x$ for $x\in[0,1]$;
\item the generalized mean method (\cite{VW20}): $\phi(x)=x^r$ for $x\in[ 0,1] $, where $r\in[-\infty,\infty]$  (for $r\in \{-\infty, 0,\infty\}$, the averaging function is obtained by taking limits, and this includes the classic Bonferroni correction with $r=-\infty$);
\item  the Cauchy combination method (\cite{LX20}): $\phi(x)=\tan\left(\pi\left(x-\frac{1}{2}\right)\right)$ for $x\in[0,1]$.
\end{enumerate}
Fisher's combination method and the generalized mean method with $r\ge 0$ utilize light-tailed distributions (the exponential and beta distributions) to transform p-variables. In contrast, heavy-tailed transformations (the Pareto and Cauchy distributions) are used by the harmonic mean method, the Cauchy combination method, and the generalized mean method with $r<0$.

For different choices of combining functions  $M^{\mathbf w}_{\phi}$ and dependence assumptions of p-variables, a threshold function $g$ is assigned such that the test is valid; that is $\p( M^{\mathbf w}_{\phi}(U_1,\dots,U_n)\le g(p))\le p$, $p\in(0,1)$; see \cite{CLTW23} for thresholds of the above methods given different dependence assumptions. However, the dependence assumption of p-variables is very difficult to verify and the validity of a test procedure cannot be guaranteed if there is any model misspecification on the dependence assumption. For instance, Fisher's combination method assumes p-values are independent and its type-I error rate can explode even if the model misspecification is small. Recently, the harmonic mean and the Cauchy combination methods have received much interest due to their nice properties against model misspecification on dependence assumptions. Coincidentally, both methods use infinite-mean distributions to transform p-variables and their performances are similar and closely connected to the famous Simes method (\cite{S86}) in several senses; see \cite{CLTW23} for the comparisons of the three methods. Other combining methods based on heavy-tailed transformations are studied by, e.g., \cite{GJW23} and \cite{fang2021heavy}.
It is by now clear that transformations to an extremely heavy-tailed distribution have various advantages over other choices, making the methods of \cite{W19} and \cite{LX20} quite popular. Under independence or some other nice dependence models, the threshold $g$ for the Cauchy combination method can be chosen as precisely $g(p)=p$, $p\in(0,1)$, and the one with the harmonic mean often needs to be corrected substantially; see \cite{CWWZ24}.
These studies show how infinite-mean models are useful in statistics even when the underlying problem does not have such a feature. 

% Due to the asymptotic properties of the harmonic mean method, one may roughly choose $g(p)=p$, $p\in(0,1)$, as its threshold under various dependence assumptions including independence. However, \cite{CWWZ24} showed that the harmonic mean method is not valid with such a threshold given several dependence assumptions including independence, a form of negative dependence, and a type of positive dependence modelled by Clayton copulas; some of the results are built on \eqref{eq:*}. In particular, if p-variables are independent, the type-I error rate of the harmonic mean method with threshold $g(p)=p$, $p\in(0,1)$, will explode as the number of p-variables increases, whereas the Cauchy combination and the Simes methods with the same threshold are always valid for any number of independent p-variables. %Moreover, the Simes method remains valid for test statistics following a multivariate normal distribution with nonnegative correlations (\cite{sarkar1998some}) and so does the Cauchy combination method for p-variables with some other dependence structures. Thus to keep the advantages of the harmonic method without correcting the threshold $g$, one may consider using the Simes and the Cauchy combination methods instead.

\section{Infinite-mean models in some other settings}
\label{sec:other}

We briefly discuss, in a few different contexts, 
how  infinite-mean models lead to different phenomena, compared  to usual finite-mean models.
In each of these settings, the corresponding conclusion flips when switching from finite-mean models to infinite-mean models. 

\subsection{Optimal bundling}\label{sec:10}

Product bundling is a strategy to sell goods as collections. It turns out that the valuation distributions may play an important role in determining the optimal bundling strategies. 
% The optimal bundling strategy may depend on whether the valuations of goods are extremely heavy-tailed or not. 
We briefly describe the Vickrey auction model studied by \cite{P83} below. 
Consider a single seller who provides $m$ goods to $n$ buyers. Let $2^M$ be the power set of $M=[m]:=\{1,\dots,m\}$. A \emph{bundling decision} $\beta=\{B_1,\dots,B_l\}$ is a partition of $[m]$.
% such that 
% \begin{enumerate}[(a)]
% \item 
% $B_s\neq\emptyset$ for any $s\in[l]$,
% \item 
% $B_s\cap B_t=\emptyset$, for any distinct $s,t\in[l]$
% \item
% $\bigcup_{s=1}^lB_s=[m]$.
% \end{enumerate}
For $j\in[n]$, the $j$th buyer's valuations for goods are represented by $\mathbf{X}_j=(X_{1j},\dots,X_{mj})$ where $X_{ij}$ is their valuation for good  $i\in[m]$ and $X_{1j},\dots,X_{mj}$ are iid. The valuations of buyers are independent of each other, i.e., $\mathbf X_{1},\dots, \mathbf X_{n}$ are independent. Each buyer does not know the valuations of the other buyers; the only available information is their own valuations and the distributions of the other buyers' valuations. For a bundle $B\in 2^M$, its  valuation of buyer $j$ is $v_j(B)=\sum_{i\in B}X_{ij}$.
 In the Vickrey auction,  all buyers will simultaneously submit sealed bids for bundles of goods.  The buyer with the highest bid wins the goods and pays the second highest bid. Buyer $j\in[n]$ with realized valuations $\mathbf{X}_j=\mathbf{x}_j\in\R^m$ has the following expected surplus for bundle $B\in 2^M$
\begin{align*}
\E[S_j(B,\mathbf{x}_j)]=\p\(\max_{s\in [n]/\{j\}} v_s(B)<v_j(B)\)
 \(v_j(B)-\E\(\max_{s\in [n]/\{j\}} v_s(B)|\max_{s\in [n]/\{j\}} v_s(B)<v_j(B) \)\).
\end{align*}
The expected surplus for a bundling decision is the sum of the expected surplus for each bundle. A buyer prefers a bundling decision if its expected surplus is larger.

In a market of two buyers, \cite{P83} showed that separate auctions ($\beta=\{\{1\},\dots,\{m\}\}$, i.e., one good is sold at a time) are preferred by buyers regardless of the valuation distributions. In the case of more than two buyers, the optimal bundling to the buyers depends on the valuation distributions. If the valuations are bounded, \cite{P83} showed that the buyers never prefer separate auctions. However, if the valuations are positive stable random variables with infinite mean,  separate auctions are preferred, as shown in Theorem 4.1 of \cite{IW10}.

\subsection{Firm growth analysis}\label{sec:11}

\cite{I14} reconsidered the demand-driven innovation and spatial competition overtime model of \cite{JR87} in the presence of extremely heavy-tailed customer signals. We briefly describe below the setting studied by \cite{I14}, a simplified version of the original model.

In each period, a firm has two decisions to make. In period $t$, the products of the firm are differentiated by a parameter 
$\hat \theta_t\in\R_+$. There exists an ``ideal" product $\theta_t\in\R_+$ but this information is not available to the firm until they decide $\hat\theta_t$. Once the firm commits to $\hat \theta_t$ and learns the ``ideal" product $\theta_t$, the price of the product, denoted by $p(\hat \theta_t,\theta_t)$, is determined. Then, the firm needs to choose the level of output $y_t$. After that, the firm will receive customers' signals about next period's ``ideal" product  $s_{i,t}=\theta_{t+1}+\epsilon_{i,t}$, $i=1,\dots,N$, where $\epsilon_{i,t}$, $i=1,\dots,N$, are iid shocks and  $N$ is a random sample size with a known distribution given $y_t$. The signals will be used to estimate next period's ``ideal" product $\theta_{t+1}$, via, e.g., the sample mean of the signals $\hat\theta_{t+1}=\sum_{i=1}^Ns_{i,t}/N=\theta_{t+1}+\sum_{i=1}^N\epsilon_{i,t}/N$. To choose $y_t$, the firm needs to solve the following problem
$$V(p(\hat \theta_t,\theta_t))=\max_{y_t}\left\{y_tp(\hat \theta_t,\theta_t)-c(y)+\beta V(x)\mathrm{d} F(x|y_t)\right\},$$
where $c$ is a convex cost function, $\beta\in\R$ is a discounting factor, and 
$F(x|y_t)$ is the distribution function of $p(\hat \theta_{t+1},\theta_{t+1})$ given $y_t$.

Let $y_t^{1}$ and $y_t^{2}$ be the outputs of two firms in period $t$. Denote by $\mathrm{RP}_t=\p(y_{t+1}^1>y_{t+1}^2|y_t^2>y_t^1)$, called the reversal rank probability.  Suppose the firms use the sample mean to estimate product design. If the signals follow a log-concave distribution, as implied by the results of \cite{JR87}, then 
$\mathrm{RP}_t\le 1/2$ and $\mathrm{RP}_t$ diminishes as $y_t^2-y_t^1 $ increases. The results also hold if the signals follow a symmetric stable distribution with a tail parameter being strictly greater than 1 (\cite{I14}). These results suggest that if customers' signals are not extremely heavy-tailed, the firms' outputs exhibit some positive persistence (i.e., firms with large outputs are likely to have large outputs in the next period). However, if the signals follow an extremely heavy-tailed symmetric stable distribution, the previous results are reversed and firms' outputs show a pattern of anti-persistence. \cite{I14} also showed that if the sample median is used as the estimator, firms' outputs are persistent regardless of whether the signals are extremely heavy-tailed or not.

\subsection{Efficiency of linear estimators}\label{sec:12}

Consider the model 
$$X_i=\mu+\epsilon_i,~~i=1,2,\dots,$$
where $\mu$ is the location parameter
and $\epsilon_i$, $i=1,2,\dots$, are iid symmetric random noises with mode 0. If $\epsilon_i$, $i=1,2,\dots$, have finite mean, then $\mu$ is the population mean. A linear estimator for $\mu$ is $\hat\theta_{\mathbf a}=\sum_{i=1}^na_iX_i$ where $\mathbf a=(a_1,\dots,a_n)\in[0,1]^n$ and $\sum_{i=1}^na_i=1$. We write $\hat\theta_n:=\hat\theta_{\mathbf a}$ for  $ \mathbf a=(1/n,\dots,1/n)$. If $\epsilon_i$, $i=1,2,\dots$, have finite second moments, it is common to use variance to compare two unbiased estimators. However, this approach fails for heavy-tailed noises. We only present here one possible solution that is the most relevant to our discussion. A formal treatment of robust statistical methods can be found in \cite{HR09}.

To solve this issue, \cite{I07} proposed to compare estimators via peakedness. For two random variables $X$ and $Y$, $X$ is said to be more \emph{strictly peaked} about $\theta\in\R$ than $Y$ if $\p(|X-\theta|>\epsilon)< \p(|Y-\theta|>\epsilon)$ for all $\epsilon>0$. An estimator $\hat \theta^1$ is said to be more \emph{$P$-efficient} than another estimator $\hat \theta^2$ if $\hat \theta^1$ is strictly more peaked about $\theta$ than $\hat \theta^2$. 
If $\hat\theta_n$ becomes more $P$-efficient as $n$ increases,  $\hat\theta_n$ is said to exhibit \emph{monotone consistency}.
\cite{I07} considered the case when $\epsilon_i$, $i=1,2,\dots$, follow a symmetric stable distribution. If the stable distribution has finite mean, it is shown that $\hat\theta_n$ is the most  $P$-efficient linear estimator and $\hat \theta_n$ exhibits monotone consistency.  However, if the stable distribution does not have finite mean,   then $\hat\theta_n$ is the least $P$-efficient linear estimator and the $P$-efficiency of $\hat\theta_n$ decreases as $n$ increases. These results hold for more general classes of noise distributions in  \cite{I07}.

\section{Conclusion}\label{sec:conclusion}

% [Warn the use of usual methods in the presence or possibility of infinite-mean models]

Heavy-tailed data, prevalent in a wide range of applications, can violate the assumption of a finite first moment, which is the theoretical foundation of many classic results such as the central limit theorem. This paper reviews a few recent advancements for infinite-mean models and discusses the breakdown of standard thinking due to infinite mean in several research areas including but not limited to portfolio diversification, risk sharing, and multiple hypothesis testing. Most of the reviewed results for infinite-mean models show a significant divergence compared to their counterparts in finite-mean cases, making it necessary to rethink the impacts of traditional methods in the presence of infinite mean. This is illustrated by many examples in this paper. Moreover, many classic methods for finite-mean models are warned to be inappropriate for infinite-mean models. For instance, the regulatory risk measure Expected Shortfall may yield infinite value when applied to infinite-mean losses, making its meaning as a capital reserve difficult to interpret. Consequently, in the presence of infinite-mean models, decision makers need to take a step back and carefully reconsider the possible consequences of applying classic methods developed in a finite-mean world.

\subsection*{Acknowledgements}
The authors thank an Editor, an Associate Editor, and two anonymous referees for helpful comments on the paper.

{

}

\end{document}